\documentstyle[preprint,aps]{revtex}
\begin{document}

\title{A Possible Einstein--Podolsky--Rosen Probe \\ of the
Momentum--Position Uncertainty Relation} \author{Roy Ringo$^*$}
\address{\underline{Physics Division, Argonne National Laboratory,
Argonne, IL 60439}}

\maketitle \begin{abstract}

It is suggested that a measurement of the products of photoemission by
alkali atoms excited after extraction from a trap, might, using the
EPR strategy, show a significant violation of the momentum-position
uncertainty relation.  If this failed, as is quite likely, possible
causes, such as retroactive propagation of influences and retrodiction
failure, could be tested on the proposed apparatus.  \end{abstract}

\bigskip \bigskip \bigskip \noindent Keywords: momentum-position
uncertainty, photoemission, trapped atoms, \\ \indent\indent\indent
\,\,\,\,                         retroactive influences \\

\bigskip \bigskip \bigskip \noindent *Electronic address:
ringo@anlphy.phy.anl.gov;  FAX: (630) 252-6210

\newpage \section{Introduction} It seems generally believed that the
Einstein, Podolsky and Rosen (EPR) \cite{einstein35} proof of the
incompleteness of quantum mechanics is wrong because it involved the
assumption of locality i.e., causes cannot propagate faster than light
\cite{mermin90,greenberger90}.  A number of experiments, notably those
of Aspect and his colleagues \cite{aspect82}, appear to show that nature
violates the limits of Bell's theorem \cite{bell65} and this can only
be understood if some details of the detection event at one detector
are available at the coincidental detection event at another detector,
although the two events have a space-like relation.  Unless one abandons
the idea of explanation and settles for descriptions of nature, this
implies the transmission of influences faster than light.  Since the
EPR proof assumed influences could be isolated, it is apparently wrong.
``Apparently" because the proof of the violation of Bell's theorem is
not quite complete \cite{ferrero90}.

Most of these experiments were done using correlations of spins.  In this
circumstance it seems reasonable to test whether these conclusions apply
to other complementary observable pairs and proposals to do this have
been made \cite{zukowski91,davis89,franson90}.  It might be particularly
interesting to look at the complementary pair EPR considered, i.e.,
momentum and position.  Several suggestions have been made for experiments
to do this.  Maric, Popper and Vigier \cite{maric88} proposed to look
at the transverse momenta and positions of annihilation gamma rays.
The experiment though did not have adequate control of the momentum of
the annihilation pair so it would only show a $\Delta p\Delta x$ product
substantially less than $\hbar /2$ in the past.  However, it has been
generally thought since the earliest days of the uncertainty principle
that in retrodiction $\Delta p \Delta x$ can be much less than $\hbar /2$.
Heisenberg \cite{heisenberg30} agreed but argued that such calculations
do not lead to any predictions so are untestable.  Scheer {\it et\,al.}
\cite{scheer89} have pointed out that measurements by Nuttall and Gallon
\cite{nuttall76} give a clear example of such a retrodiction, taking
advantage of a monatomic layer of argon to get a small $\Delta x$.
However, the interaction of every atom with its substate may seriously
complicate such experiments.

This suggests it may be best to do such experiments on an isolated excited
atom, measuring the momentum of the photon it would emit very precisely
and predicting the trajectory of the recoiling atoms very precisely using
the Einstein-Podolsky-Rosen scheme.  The recent development of traps
for atoms may make such an experiment possible although far from easy.

A quantum mechanical prediction of the dispersion product for this
experiment would give a product $\geq \hbar /2$.  The experiment
however is intended as a test of the completeness of quantum mechanics.
If quantum mechanics is incomplete, there may be special circumstances
where $\Delta p\Delta x << \hbar /2$.  To even examine the possibility
of a much smaller product, a different analysis must be used.  There is
little guidance as to what that should be.  For the purposes of this
paper, an analysis will be done using the following assumptions; first,
strict conservation of energy and momentum; second, no retroactive
influence; third, precise knowledge of position and momentum in the
past (retrodiction) \cite{heisenberg30}; fourth, the semiclassical
approximation for the calculation of trajectories \cite{heller93}.
Tests of these assumptions, which may well be the most useful product
of this proposal, are discussed in some detail later.

\section{Estimate of the Dispersion Product}

The general stategy of the proposed experiment is to produce an ensemble
of objects whose position and momentum can, using conservation of
momentum, be predicted to an accuracy such that the dispersion product
of the individual objects would be less than $\hbar$/2 and then to test
these predictions by measuring the trajectories of these objects with
great accuracy.

The experiment could proceed as follows: atoms in a beam of high phase
space density, ejected from a trap, are incident in the z direction on a
slit, narrow in the x direction, which delimits their x position (Fig. 1).
(In these measurements the horizontal x-dimension is the crucial one.
The effects of motion in the other dimensions are fairly obvious.)
Before passing through the slit they encounter an intense light beam in
the y direction.  The light is of the right wave length to excite the
atom to a state from which it decays in the order of a $\mu s$.  Protons
in the -x direction are selected by a high resolution spectrometer.
This selection process should make it possible to predict the trajectory
of the atom in delayed coincidence with the spectrometer detector using
conservation of momentum.  The accuracy of the prediction can be tested
by the precision, $\Delta t_a$, with which it predicts the flight time
of the atom at the atom detector.  ($t_a$ is the flight time from the
emission of the photon to the atoms arrival at the detector.)

The calculation of the dispersion of the flight time uses the assemptions
listed.  It starts with the estimation of the dispersion of the initial
time, $\Delta t_i$.  This is essentially $1/2$ the duration of the
exciting light pulse plus 1/2 the decay time of the excited state.
This should be no greater than $10^{-5}$s at worst.  Almost all the
uncertainty in the prediction of $t_2$ comes from the trajectory of the
atom after emission.  The velocity dispersion is dominated by that due
to diffraction at the slit.  The dispersion in the arrival time at the
atom detector can be written as

\begin{equation} \Delta t_a  = \Big[ (\Delta t_i )^2 + (\Delta x_s/v_a
)^2 + (x\Delta v_a /v^2_a\,) \Big] ^{1/2} \,.  \end{equation} Here $x$
is the distance from the slit to the atom detector, $v_a$ is the atoms
x-velocity after going through the slit and emitting the photon, and
$\Delta$ means the standard deviation or a reasonable approximation to
it, e.g.\ $\Delta x_s$ is taken as 1/4 the slit width.  $\Delta x$ is a
rather complicated matter and the discussion of that will be postponed
until some parameters of the apparatus have been discussed.  For the
present purposes, it is enough to say that the term containing $\Delta
x_s$ is suffiently smaller than the last term that it can be neglected
as can $\Delta t_i$.

The measurement of the atom x-velocity is critical.  The photon energy is
shifted from that of a motionless free atom by the Doppler effect from
the motion of the atom.  The spectrometer needs a resolution sufficient
to give the atoms x-velocity with an accuracy of about 1/30 of the
diffraction-induced motion at the slit.  From Eq.\ (1) we can deduce

\begin{equation} \Delta v_a = {v^2_a \Delta t_a \over x} \,.
\end{equation} This combined with the $\Delta x$ to be discussed can
then be compared with the uncertainty principle value of the dispersion
product $\hbar /2$

\section{Apparatus}

Figure 1 is a sketch of a conceivable arrangement of the apparatus.
Illustrative parameters of the appartus are shown in Table I.
One critical feature is an atomic trap that will give phase space
densities of at least $10^{17}$ atoms cm$^{-3}$ (m/s)$^{-3}$.  While this
is possible, it is not simple \cite{bradley95}.  Moreover, there is a real
problem in ejecting atoms from the trap which will probably operate in a
pulsed mode.  Ejection should be by the method that best preserves the
phase space density and maximizes (density) x (frequency of ejection).
However, these densities are high enough that, if they can be exploited,
counting rates should not be a fatal problem.

The atoms in the trap should be alkali, since there is considerable
experience in trapping them and even neutral alkali atoms are relatively
easy to detect.  After extraction from the trap, the atoms would be
radiated by a laser of the appropriate frequency and power to lift them to
a state of a lifetime of at least 0.1 $\mu s$.  The demands on the laser
are not severe. It needs about 1 watt of power focussed on 10$^{-12}$
m$^2$.  It would be pulsed in near synchronism with the ejection of
atoms from the trap, for 10$^{-5}$ seconds, perhaps 100 times a second.
The focusing of the laser beam will give the photons small components in
the x-direction which will be transformed to the momentum of the atoms.
If there is a $\Delta\theta _x$ = 0.05 radians spread in the x-direction,
the velocity spread induced in the atoms would be

\begin{equation} \Delta v_{exit} = {\Delta\theta _x \cdot E\,(\rm{photon})
\over c \cdot m} = 0.6 \times 10^{-2}\,\, $m/s$ \,.  \end{equation} This
is negligible compared to the dispersion from diffraction in the slit.

The neutral atoms would be detected at $t_a$ by a surface ionization
detector, essentially a flat plate  of hot rhenium perpendicular to the
x-direction to within 10$^{-6}$ rad.  When the atoms are ionized they are
attracted to a microchannel plate.  Because the intrinsic z-velocity of
the atoms from the trap should be $\sim$ 10 m/s they would be well away
from the slit in the z direction when detected.  This should make it
relatively easy to move the ions, in a uniform way, by an electric field
pulling them into the microchannel plate with negligible variance in the
time delay for all the atoms involved in the experiment.  Because the
velocity of the accelerated ions is relatively high, the flatness and
alignment of the microchannel array is not of major concern.  The channels
would all be connected in parallel.

$^7 Li$ appears to be the best alkali to use because it minimizes the
needed resolution of the spectrometer.  The desorption time is of some
concern here but with rhenium at 1400 K, $Li$ should desorb in $<10^{-2}$
s \cite{scheer63}.  It would be of a modest help to use separated $^7
Li$ but since $^6 Li$ is only 7.5\% of $Li$, it would at worst be a
small addition to background.  A suitable spectrum line from $Li$ is
(1s)$^2$ $3p^2 P \rightarrow$ ground state (323 nm and a lifetime of 0.8
$\times$ 10$^{-6}$ s).  This would give a recoil velocity of 0.12 m/s.
The lifetime gives a spread of $\Delta E/E$ ($E$ is the photon energy)
of 1.4 $\times$ 10$^{-10}$.  This limits the useful resolution of the
spectrometer to 7 $\times$ 10$^9$.

The dispersion in $t_a$ is increased by the motion of the atom after it
passes the slit.  The dispersion in the x-velocity of the atom $\Delta
v_A$ will lead to a dispersion in x of about $t_a \Delta v_A$, where $t_a$
is the time after the atom passes the slit.  The spectral line used in
this experiment must be fairly narrow or $\Delta E(=\hbar \Delta t)$ will
be larger than the spread coming from the doppler shift of the x motion.
This means excited state lifetimes of $\geq$ 10$^{-6}$ s.  Such times
would lead to an unacceptably large $\Delta x$ (from the diffraction
induced velocity).  This can be remedied by putting the exciting beam,
which moves in the - direction, between the atom trap and the slit and
screening all but a few microns past the slit from the spectrometer.
The consequent spread in x can thus be held to less than the slit width.
There is obviously a cost for this in counting rate but it appears to
be necessary.  The x-dispersion from the slit shown in Fig.\ 1 should be
about 0.5 $\times$ 10$^{-8}$m.  The dispersion of the x-position of the
detector plate is taken as the same.  These two folded together and with
the $\Delta x$ from the diffractive spreading give a final $\Delta x$
of 1.0 $\times$ 10$^{-8}$m.  This width of the defining slit is very
narrow and there is possible danger of it being narrowed further by
accummulation of $Li$ atoms.  This could be inhibited by raising the
temperatures of the slit.

The spectrometer with which one infers the recoil x-velocity of the
atoms by the doppler shift in the photons they emitted must have a
very high resolution.  In the illustrative example being given $\lambda
/\Delta \lambda$ is taken as 7 $\times$ 10$^9$.  There are Fabry-Perot
cavity spectrometers of this resolution commercially available [Newport
Corporation Super-Cavity Optical Spectrum Analyzers], but not yet for
wave lengths as short as 323 nm.  Achieving the desired resolution
will surely be one of the most difficult parts of this experiment.
It should be pointed out that a possible solution to this problem is a
spectrometer using a considerably longer Fabry-Perot cavity than the
commercial one (2.5 cm).  This would need fewer reflections than the
shorter one, and might be practical in the near UV.  Alternatively,
this long cavity might give a resolution in the visible high enough to
make it possible to use a visible line from sodium.\cite{demtrotter82}

The parameters of the experiment then are given in Table I.  Given these
parameters $\Delta v_a$ from Eq.\ (2) is 5 $\times$ 10$^{-2}$ m/s, the
mass of the $^7Li$ atom is 1.6 $\times$ 10$^{-26}$ kg, then $\Delta p_a
\Delta x \approx \hbar /17$.

\section{Tests of the Assumptions}

The conservation of momentum in atom-photon interactions has been
tested well enough for the suggested dispersion product measurement
(about $m_{Li}\, 10^{-3}$ m/s) by an experiment done by Weiss, Young
and Chu \cite{weiss93}.  They use atomic interferometry in an experiment
where the ultimate aim was a precision measurement of the fine structure
constant and they show momentum conservation to $m_{C_s}$ 10$^{-7}$
m/s in photon interactions with $C_s$ atoms.

The combination of retroactive influences and failure of retrodiction,
i.e.\ the failure of the second and third assumptions could lead to
an arbitrary increase in the dispersion product.  For example, if the
detection time of the photon detector were of the order of picoseconds
and the time-energy uncertainty principle applied, then the resolution
of the photon spectrometer would be badly degraded, i.e.\ a failure of
precise retrodiction.  If retroactive influence were allowed, this in
turn could affect the selection of the momentum of the atom and that
could produce a dispersion product $\geq \hbar /2$.  Note that it would
take a failure of both assumptions to do this.  The validity of this
analysis could be tested by dispersing the detector times which could
sharpen the energy resolution.

If the preceding two assumptions pass the test, the last assumption,
the semiclassical approximation, may be inapplicable.  This is almost
equivalent to saying that the uncertainty relation is an absolute and
at least at the present stage of knowledge is not to be questioned, but
simply accepted as a foundation of physics.  This is probably right but
there is some obligation to continue testing foundations.  This is not,
however, work for people concerned with tenure.

\section{Counting Rates}

The Bose-Enstein condensation experiments give very high phase
space densities, 10$^{17}$ atoms cm$^{-3}$ (m/s)$^{-3}$ and higher.
Specifically C.C. Bradley {\it et\,al.} \cite{bradley95} appear to have
a phase space density of 2 $\times$ 10$^{17}$ $Li$ atoms, cm$^{-3}$
(m/s)$^{-3}$ at a temperature of 100 nK.  In the experiment being
described in the present paper, the geometric volume per second passing
would be given by the slit width in x, 10$^{-6}$ cm, the slit length
in y, 10$^{-2}$ cm and the beam length in 1 sec for the z dimension.
This last could be 100 pulses/s $\times$ 10$^{-2}$ cm ($\sim$ diameter
of the atomic cloud in the trap), i.e. 1 cm.  This gives a volume of 2
$\times$ 10$^{-8}$ cm$^3$.  The velocity corresponding to $Li$ at 100
nK is 1.1 $\times$ 10$^{-2}$ m/s and is fully usable in the y and z
directions, but only 1/30 is usable in the x direction because of the
diffraction in that direction, so only 3 $\times$ 10$^{-6}$ (m/s)$^3$
velocity space is filled.

There is a loss from spreading of the atomic beam from the trap to
the slit estimated to take 0.02 seconds.  The reduction of the density
should be about a factor of 3.  The spectrometer accepts only $\sim 3
\times 10^{-4}$ of the sphere of photons emitted\footnote{The effect
of this on the resolution in $V_x$ is negligible.  The maximum angle
from the x axis is 2.7 $\times$ 10$^{-2}$ and the loss in velocity, this
causes 3.1 $\times$ 10$^{-4}$ m/s which is to be compared with $<\Delta
v_x >$ = 5 $\times$ 10$^{-2}$ m/s.} and might well lose a factor of 3
or more in passing through it.  The selection of a narrow part of the
decay curve could lose about a factor of 10.  The atom drifts about
one lifetime (0.8 $\mu$sec $\approx$ 10$^{-5}$ m/10 m/s) and then is
open toward the spectrometer for 2.10$^{-6}$ m which corresponds to 0.2
lifetimes. Inefficiences in detectors and miscellaneous losses could be
another factor of 10.  Inefficiences in the excitation process and the
two detectors could be still another factor of 10.  All this leads to
about 1/10 photon counted/min.

This could be adequate if the background were small.  It should be.
The most serious source of background should be scattering of light from
the exciting laser on stationary objects.  This would however be a very
sharp line which could be avoided in the spectrometer by setting the
spectrometer at a line corresponding to, say a 0.5 m/s Doppler shift.
Moreover, scattering in general would not be in coincidence with the
detection of an atom.

This is clearly a very difficult experiment but the difficulties are
in technical matters not in fundamental physics and should lessen with
time given the role of such active fields as atom trapping and laser
spectroscopy.

\section*{Acknowledgments}

I am very grateful for useful advice provided by Paul Benioff, Joseph
Berkowitz, Gordon Berry, William Childs, Patricia Dehmer, Melvin
Freedman, Gordon Goodman, Mitio Inokuti, Thomas LeBrun, Gerald Marsh,
William Daniel Phillips, Abner Shimony, Lee Teng, and Linda Young.

This work is supported by the U.S.\ Department of Energy, Nuclear Physics
Division, under contract W-31-109-ENG-38.

\begin{table} 
\caption{}
\begin{tabular}{c} %\begin{center} %\hline
%\vspace*{0.2in} 
x = 0.01 m \\ 
$\Delta x_s$ = 0.5 $\times$ 10$^{-8}$ m \\ 
$\Delta x$ = 10$^{-8}$ m \\ 
E(photon) = 4.2 $\times$ 10$^{-19}$ J (323 nm) \\ 
R = 7 $\times$ 10$^9$   \\ 
$<v_a >$ = 1 m/s \\ $\Delta v_a$ = .05 m/s \\ 
$\Delta t_a$ = 5 $\times$ 10$^{-4}$ s \\ 
$t_a$ = 0.01 s \\
%\vspace*{0.2in} %\hline %\end{center} 
\end{tabular} 
\end{table}

\newpage

\begin{figure} \caption{Schematic of the apparatus.} \end{figure}

%\begin{figure} %\caption{Enlarged view of the center of the epxeriment.}
%\end{figure}

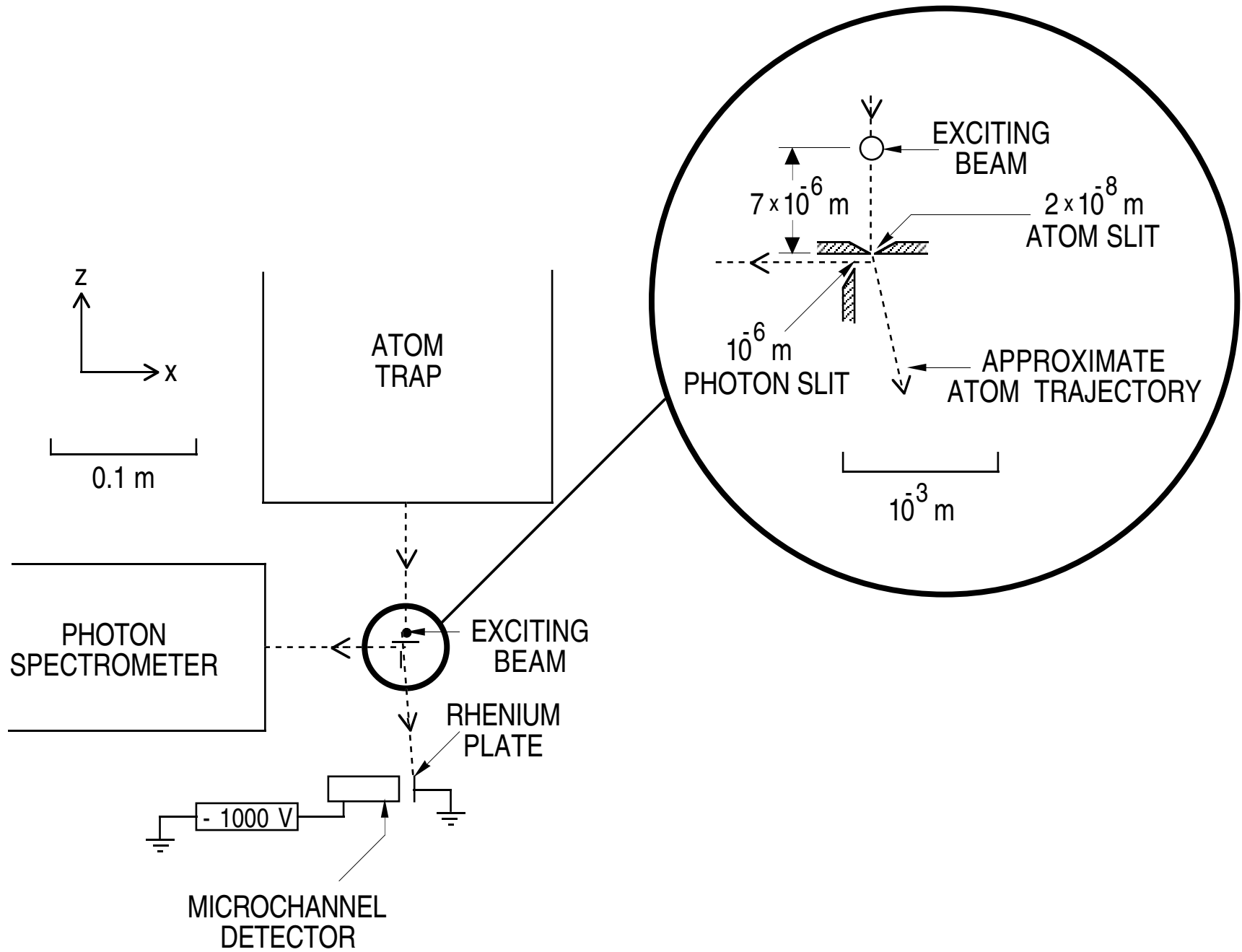
\end{document}